\documentclass[journal,a4paper]{IEEEtran}

\usepackage{import}
\usepackage{graphicx}
\usepackage{xcolor}
\usepackage{soul}
\usepackage{times}
\usepackage{epsfig}
\usepackage{enumerate}
\usepackage{amsfonts}
\usepackage{amssymb}
\usepackage{colortbl}
\usepackage{upgreek}
\usepackage{booktabs}
\usepackage{bbm}
\usepackage{multicol}
\usepackage{lipsum}
\usepackage{cite}
\usepackage{algorithm}
\usepackage{algpseudocode}
\usepackage[acronym, nonumberlist]{glossaries}
\usepackage[automake]{glossaries-extra}
\usepackage{amsmath}
\usepackage[cmintegrals]{newtxmath}
\usepackage{bm}
\usepackage{url}
\usepackage[caption=false,font=footnotesize]{subfig}
\usepackage{multirow}
\usepackage{array}
\usepackage{pifont}
\usepackage{comment}
\usepackage{circledsteps}
\usepackage{tikz}
\usetikzlibrary{shapes.geometric, arrows}
\usepackage{dsfont}
\usepackage{bbold}

\ifCLASSINFOpdf
\else
\fi
\newglossarystyle{mystyle}{%
  \setglossarystyle{long}%
     {\begin{longtable}{p{1.0cm}p{\glsdescwidth}}}%
     {\end{longtable}}%
}
\newglossarystyle{modsuper}{%
  \setglossarystyle{super}%
}

\newglossaryentry{LV}{name={LV}, description={Low Voltage}}
\newglossaryentry{MV}{name={MV}, description={Medium Voltage}}
\newglossaryentry{GIS}{name={GIS}, description={Geographic Information System}}
\newglossaryentry{SI}{name={SI}, description={System Identification}}
\newglossaryentry{TI}{name={TI}, description={Topology Identification}}
\newglossaryentry{PI}{name={PI}, description={Parameter Identification}}
\newglossaryentry{PHI}{name={PHI}, description={Phase Identification}}
\newglossaryentry{PMU}{name={PMU}, description={Phasor Measurement Unit}}
\newglossaryentry{NPMU}{name={NPMU}, description={Non-Phasor Measurement Unit}}
\newglossaryentry{BLP}{name={BLP}, description={Binary Linear Program}}
\newglossaryentry{WLS}{name={WLS}, description={Weighted Least Squares}}
\newglossaryentry{MIQP}{name={MIQP}, description={Mixed-Integer Quadratic Program}}
\newglossaryentry{GPS}{name={GPS}, description={Global Positioning System}}
\newglossaryentry{NTP}{name={NTP}, description={Network Time Protocol}}
\newglossaryentry{AC}{name={AC}, description={Accuracy Class}}
\makeglossaries

\begin{document}
	
	\title{Data-driven Topology and Parameter Identification in Distribution Systems with limited Measurements}	
	
	\author{
		Steven de Jongh, \IEEEmembership{Student Member,~IEEE}, 
		Felicitas Mueller, \IEEEmembership{Student Member,~IEEE}, 
		Fabian Osterberg, \newline Claudio A. Ca\~{n}izares, \IEEEmembership{Fellow,~IEEE}, 
        Thomas Leibfried,~\IEEEmembership{Senior Member,~IEEE} and
        Kankar Bhattacharya, \IEEEmembership{Fellow,~IEEE} 
		\thanks{S. de Jongh, F. Mueller, F. Osterberg and T. Leibfried are with the Institute of Electrical Energy Systems and High Voltage Engineering (IEH) of the Karlsruhe Institute of Technology (KIT), Karlsruhe, Germany. \newline \indent C. Ca\~{n}izares and K. Bhattacharya are with the Department of
        Electrical and Computer Engineering at the University of Waterloo, ON,
        Canada. \newline \indent 
        This work is funded by MITACS, Karlsruhe Institute of Technology and University of Waterloo.
        \newline \indent The A.I. based language models DeepL, OpenAI GPT-3.5 and you.com were used for text editing and other language-related tasks.}
	}
	
	
	
	\maketitle 
	\begin{abstract}
      This manuscript presents novel techniques for identifying the switch states, phase identification, and estimation of equipment parameters in multi-phase low voltage electrical grids, which is a major challenge in long-standing German low voltage grids that lack observability and are heavily impacted by modelling errors.  The proposed methods are tailored for systems with a limited number of spatially distributed measuring devices, which measure voltage magnitudes at specific nodes and some line current magnitudes. The overall approach employs a problem decomposition strategy to divide the problem into smaller subproblems, which are addressed independently. The techniques for identifying switch states and system phases are based on heuristics and a binary optimization problem using correlation analysis of the measured time series. The estimation of equipment parameters is achieved through a data-driven regression approach and by an optimization problem, and the identification of cable types is solved using a Mixed-Integer Quadratic Programming solver. To validate the presented methods, a realistic grid is used and the presented techniques are evaluated for their resilience to data quality and time resolution, discussing the limitations of the proposed methods.
	\end{abstract}
	
	\begin{IEEEkeywords}
        Low voltage grids, parameter estimation, phase identification, state estimation, system identification, topology identification.
    \end{IEEEkeywords}
	
	\ifCLASSOPTIONpeerreview
	\begin{center} \bfseries EDICS Category: 3-BBND \end{center}		
	\fi
	\IEEEpeerreviewmaketitle											

    \section*{Notation }
    The following notation is used throughout this paper: Caligraphic letters are used to denote sets or distributions (e.g., $\mathcal{A}$). Scalars and single dimensional variables are denoted using small letters (e.g. $a$), whereas capital and bold letters represents matrices (e.g. $\mathbf{A}$), and lower case letters vectors (e.g., $\mathbf{a}$). Underlined variables are used for phasor quantities (e.g., $\underline{a}$), $\mathbb{1}_{n \times m}$ is used to describe a matrix of dimension $n \times m$ containing only ones, $\mathds{1}_n$ stands for the identity matrix of dimension $n \times n$, and $\mathbb{0}_{n\times m}$ is a matrix of dimension $n \times m$ containing only zeros. The $\hat{\cdot}$ notation is used to refer to estimated values and $\otimes$ and $\cdot$ denote the Kronecker product and Matrix-vector product, respectively. Superscript ones (e.g. $a^1$) are used to refer to the positive sequence component of $a$.
    \newpage
    \noindent The following acronyms are used throughout the paper:
    \printglossary[title=, style=modsuper]

	\section{Introduction}
	\IEEEPARstart{T}{he} availability of accurate electrical models of distribution grids is increasingly important due to the integration of new types of loads and generation, such as heat pumps, electric vehicles, and photovoltaic generators at Low Voltage (\gls{LV}) and Medium Voltage (\gls{MV}) level. This is particularly true for long-standing \gls{LV} networks for which little information is available, as in the case of some German \gls{LV} grids. As the energy transition accelerates, it is expected that the additional stress caused by new grid elements will increase. In this context, accurate physical models of the system are essential for calculating flows in the grid and estimating grid load caused by emerging assets, as well as for new operational management approaches. However, the most common method of creating models of non-active distribution grids, based on data from Geographic Information Systems (\gls{GIS}), can lead to errors and inconsistencies due to data quality issues and misinterpretations in the conversion process. An alternative approach is to use time series of grid measurements in combination with data-driven methods to apply System Identification (\gls{SI}) techniques. Thus, this paper investigates the potential of \gls{SI} methods for model generation in distribution grids, with a focus on a limited number of measuring devices to represent realistic conditions, examining the robustness, data requirements, and estimation accuracy.

    There are several approaches in the literature to perform parameter and Topology Identification (\gls{TI}) for electric power systems. A detailed overview of methods for identification under different conditions is given in \cite{overview1}. The discussed methods focus either on \gls{TI} \cite{topology1, topology3, topology4} or \gls{TI} and PHase Identification (\gls{PHI}) \cite{topology2}. Methods for grid Parameter Identification (\gls{PI}),  given the true topology, are presented in \cite{param1} and \cite{param2}. In \cite{sparsereg1, patopa, patopaem}, both \gls{TI} and \gls{PI} problems are considered in a coupled fashion. When applying \gls{SI} in electrical grids, a distinction must be made between transmission and distribution grids due to the different unique preconditions, such as assumptions regarding shunts and the R/X ratio. Untransposed multi-phase distribution system \gls{SI} are considered in \cite{distribution5} and \cite{distribution1}. In \cite{distribution6}, Kron reduced systems are used to allow the estimation of all parameters where not all nodes are equipped with measurement devices. Since system phasor measurements of voltages and currents are not readily available in distribution systems, \cite{distribution7, distribution8, distribution3, distribution4} consider measurements of voltage and current magnitudes. A limited number of measured nodes is considered in \cite{distribution1}, recommending more than 80\% of measured nodes for satisfactory results.
    
    The methods used for \gls{SI} differ based on the assumptions made regarding the measurement equipment. In cases where not all nodes are measured, \cite{stats1} and \cite{stats2} apply statistical methods based on the calculation of covariances, minimum spanning trees, and Bayesian networks. For grids where measurements from all nodes are available, methods based on the underlying load flow equations are proposed in the literature. Thus, to account for noise on current and voltage measurements, optimization problems in the form of total least squares are applied, such as in \cite{inversepowerflow} and \cite{tls1}. In \cite{bayesian, bayesianEXT, OnlineLearning}, the \gls{SI} is solved using an $\ell_1$-norm optimization problem such that a sparse system matrix is obtained. In some applications, \gls{TI} is incorporated into classical state estimation in an approach known as generalized state estimation, as in \cite{gse1} and \cite{gse2}, which allows the estimation of system states and switch positions.

    The approach presented in this paper adds to the existing literature by introducing methods that apply \gls{SI} to distribution grids where the number of available measurement points is limited and substantially lower compared to previous approaches, based on measurements of only current and voltage magnitudes. Thus, various methods are presented that allow to estimate line parameters, switch states, and cable types in multi-phase \gls{LV} distribution systems under realistic assumptions with a small number of  measurements. In contrast to the approaches available in the literature, the problem is divided into different subproblems, each of which is solved using tailored approaches, which allows for a realistic modelling of the complete procedure of identification of \gls{LV} distribution grids.

    \noindent The rest of this paper is structured as follows: In Section \ref{sec:modelling}, the underlying models of the electrical system are introduced. Section \ref{sec:SImethods} presents the proposed identification procedures and the required pre-processing steps as well as the methods developed for \gls{TI}, \gls{PI}, and \gls{PHI}. The results of a simulation study are presented in Section \ref{sec:simandresults} by first introducing the test grid and related measurements in Section \ref{sec:testsystemmeasurements} and discussing the results in detail in Section \ref{sec:results}. Finally, the main conclusions of the presented work are provided in Section \ref{sec:conclusionoutlook}.

    \section{Grid Modelling}
    \label{sec:modelling}
    \newcommand{\cplx}{\mathrm{j}}
	\newcommand{\vect}[1]{\mathrm{#1}}
    An electrical distribution grid can be modeled as a graph $\mathcal{G}$ consisting of nodes $\mathcal{N}=\{1,2,\dots,n\}$ and edges $\mathcal{E}=\{(k,l), \quad k, l \in \mathcal{N}\}$. Nodes usually represent busbars at which consumers are connected or are connection points between edges of the system, which include overhead lines, cables, switches, and transformers.
    
    The voltage drop between node $k$ and $l$ in a three-phase system is defined as follows \cite{distribution5}:

    \begin{equation}
        \underbrace{\begin{bmatrix} \underline{v}_{k,\mathrm{A}} - \underline{v}_{l,\mathrm{A}} \\ \underline{v}_{k,\mathrm{B}} - \underline{v}_{l,\mathrm{B}}\\ \underline{v}_{k,\mathrm{C}} - \underline{v}_{l,\mathrm{C}} \end{bmatrix}}_{\Delta \underline{\mathbf{v}}^{\mathrm{ABC}}_{kl}} = \underbrace{\begin{bmatrix} \underline{z}_{\mathrm{A}\mathrm{A}} & \underline{z}_{\mathrm{A}\mathrm{B}} & \underline{z}_{\mathrm{A}\mathrm{C}} \\ \underline{z}_{\mathrm{B}\mathrm{A}} & \underline{z}_{\mathrm{B}\mathrm{B}} & \underline{z}_{\mathrm{B}\mathrm{C}} \\ \underline{z}_{\mathrm{C}\mathrm{A}} & \underline{z}_{\mathrm{C}\mathrm{B}} & \underline{z}_{\mathrm{C}\mathrm{C}} \end{bmatrix}}_{\mathbf{\underline{Z}}_{kl}^{\mathrm{ABC}}} \cdot \underbrace{\begin{bmatrix} \underline{i}_{kl,\mathrm{A}} \\ \underline{i}_{kl,\mathrm{B}} \\ \underline{i}_{kl,\mathrm{C}} \end{bmatrix}}_{\underline{\mathbf{i}}_{kl}^{\mathrm{ABC}}}
        \label{eq:voltagedropline}
    \end{equation}
    where $\Delta \underline{\mathbf{v}}^{\mathrm{ABC}}_{kl}$ is the phasor voltage drop between nodes $k$ and $l$ for phases $\mathcal{P} = \{ \mathrm{A},\mathrm{B}, \mathrm{C} \}$, which is obtained by multiplying the impedance matrix $\mathbf{\underline{Z}}_{kl}^{\mathrm{ABC}} \in \mathbb{C}^{3\times 3}$ and the complex line currents of all phases $\underline{i}_{kl,c} \; \forall c \in \mathcal{P}$. The impedance matrix $\mathbf{\underline{Z}}_{kl}^{\mathrm{ABC}}$ is assumed to be symmetric, which is typically the case in the networks of interest.
    
    In many applications, the measurements of voltages and currents are not available as phasors, since Phasor Measurement Units (\gls{PMU}) are necessary for these types of measurements, which are rarely deployed in \gls{LV} distribution grids. Hence, the measurements are available only as magnitudes of the complex quantities, where $\underline{v}_{k,c} = v_{k,c} \mathrm{e}^{\mathrm{j}\theta_{k,c}}, \forall k \in \mathcal{N}, c\in \mathcal{P}$, and $\underline{i}_{kl,c} = i_{kl,c} \mathrm{e}^{\mathrm{j}\phi_{kl,c}}, \forall kl \in \mathcal{E}, c \in \mathcal{P}$. In this case, the following approximation can be used:

    \begin{equation}
        \begin{bmatrix} v_{k,\mathrm{A}} - v_{l,\mathrm{A}} \\ v_{k,\mathrm{B}} - v_{l,\mathrm{B}} \\ v_{k,\mathrm{C}} - v_{l,\mathrm{C}} \end{bmatrix} \approx \begin{bmatrix} z_{\mathrm{A}\mathrm{A}} & z_{\mathrm{A}\mathrm{B}} & z_{\mathrm{A}\mathrm{C}} \\ z_{\mathrm{B}\mathrm{A}} & z_{\mathrm{B}\mathrm{B}} & z_{\mathrm{B}\mathrm{C}} \\ z_{\mathrm{C}\mathrm{A}} & z_{\mathrm{C}\mathrm{B}} & z_{\mathrm{C}\mathrm{C}} \end{bmatrix} \cdot \begin{bmatrix} i_{kl,\mathrm{A}} \\ i_{kl,\mathrm{B}} \\ i_{kl,\mathrm{C}} \end{bmatrix}
        \label{eq:voltagedroplinesimplified}
    \end{equation}
    where voltage and current angle differences are assumed to be small, i.e., $\theta_{k} - \theta_{l} \approx 0$ and $\phi_{k} - \phi_{l} \approx 0$, given the characteristics of \gls{LV} distribution grids. Furthermore, the impedance in these networks can be approximated by the magnitude $z = \sqrt{r^2 + x^2}$, with line resistance $r$ and reactance $x$, since the ratio $r/x$ in \gls{LV} grids is greater than 2, and thus the magnitude of $\underline{z}$ is dominated by its resistive component \cite{pandapower}.

    Since $\mathbf{\underline{Z}}_{kl}^{\mathrm{ABC}}$ is symmetric, symmetrical components can be defined by using the following transformation matrix:
    \begin{equation}
        \underline{\mathbf{T}}^{-1} = \frac{1}{3} \begin{bmatrix}
        1 & 1 & 1 \\ 1 & \underline{a} & \underline{a}^2 \\ 1 & \underline{a}^2 & \underline{a}
        \end{bmatrix}
        \label{eq:symmcompT}
    \end{equation}
    where $\underline{a}=\mathrm{e}^{\mathrm{j}2/3\pi}$ and positive-, negative- and zero-sequence can be obtained as follows:    
    \begin{equation}
    \begin{aligned}
        \underline{\mathbf{i}}_{kl}^{\mathrm{012}} &= \underline{\mathbf{T}}^{-1} \cdot \underline{\mathbf{i}}_{kl}^{\mathrm{ABC}} \\
        \Delta \underline{\mathbf{v}}_{kl}^{\mathrm{012}} &= \underline{\mathbf{T}}^{-1} \cdot \Delta \underline{\mathbf{v}}_{kl}^{\mathrm{ABC}} \\
        \label{eq:symmcompiv}
    \end{aligned}
    \end{equation}    
    Additionally, the matrix $\underline{\mathbf{Z}}_{kl}^{\mathrm{ABC}}$ can be transformed by using:
    \begin{equation}
        \underline{\mathbf{Z}}_{kl}^{012} = \underline{\mathbf{T}} \cdot \underline{\mathbf{Z}}^{\mathrm{ABC}}_{kl} \cdot \underline{\mathbf{T}}^{-1} = \begin{bmatrix} \underline{z}_0 & 0 & 0 \\ 0 & \underline{z}_1 & 0 \\ 0 & 0 & \underline{z}_2 \end{bmatrix}
        \label{eq:symmcompZ}
    \end{equation}
    Applying the symmetric components transformation allows the separate calculation of the system in its respective components, thus eliminating the couplings between the system phases in \eqref{eq:voltagedropline} and \eqref{eq:voltagedroplinesimplified} by not considering them explicitly.

    \section{Proposed System Identification Methods}
    \label{sec:SImethods}
    \subsection{System Identification Algorithm}
    \label{sec:systemidentificationalgorithm}

    In this section, novel methods for \gls{SI} in distribution systems are introduced. Thus, a \gls{TI} technique that leverages the estimation of switch states within the grid is presented first. The approach includes a \gls{PHI} methodology that is essential for accurately assigning the phases of individual measurements, and two \gls{PI} strategies are proposed for the estimation of total impedance and individual impedances of segments. 
    
    To enable the application of \gls{SI} methods to \gls{LV} grids with limited measured nodes, the overall problem is decomposed into smaller subproblems. Thus, a subsystem is defined as a system of electrical branch components with unknown parameters that lie between two nodes at which measurements are available. This process involves characterizing each subsystem by its measurements of currents and voltages at the input and output, as well as an arbitrary number of unmeasured nodes between them. Fig. \ref{fig:subproblem} illustrates a single-phase representation of a subsystem, where voltages at the input node $k$ and output node $l$, as well as the inflowing $\underline{i}_{\mathrm{in}}$ and outflowing current $\underline{i}_{\mathrm{out}}$ are measured. Between the two nodes $k$ and $l$ there may be an arbitrary number of grid nodes without measurements, denoted as set $\mathcal{O}$, but there are no sub-feeders.
    
    \tikzset{every picture/.style={line width=0.75pt}} 
    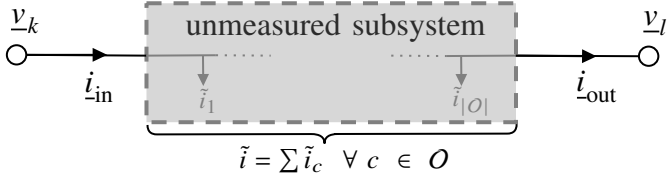
\begin{figure}[ht]
    \begin{tikzpicture}[x=0.75pt,y=0.75pt,yscale=-0.68,xscale=0.68]
    
    \draw  [color={rgb, 255:red, 128; green, 128; blue, 128 }  ,draw opacity=1 ][fill={rgb, 255:red, 155; green, 155; blue, 155 }  ,fill opacity=0.4 ][dash pattern={on 5.63pt off 4.5pt}][line width=1.5]  (189,19.35) -- (460.25,19.35) -- (460.25,107.35) -- (189,107.35) -- cycle ;
    \draw   (86.5,57.53) .. controls (86.5,53.59) and (89.69,50.4) .. (93.63,50.4) .. controls (97.56,50.4) and (100.75,53.59) .. (100.75,57.53) .. controls (100.75,61.46) and (97.56,64.65) .. (93.63,64.65) .. controls (89.69,64.65) and (86.5,61.46) .. (86.5,57.53) -- cycle ;
    \draw  [fill={rgb, 255:red, 255; green, 255; blue, 255 }  ,fill opacity=1 ] (550.5,58.53) .. controls (550.5,54.59) and (553.69,51.4) .. (557.63,51.4) .. controls (561.56,51.4) and (564.75,54.59) .. (564.75,58.53) .. controls (564.75,62.46) and (561.56,65.65) .. (557.63,65.65) .. controls (553.69,65.65) and (550.5,62.46) .. (550.5,58.53) -- cycle ;
    \draw    (100.75,57.53) -- (188.9,58.1) ;
    \draw    (100.75,57.53) -- (157.18,57.5) ;
    \draw [shift={(160.18,57.5)}, rotate = 179.98] [fill={rgb, 255:red, 0; green, 0; blue, 0 }  ][line width=0.08]  [draw opacity=0] (10.72,-5.15) -- (0,0) -- (10.72,5.15) -- cycle    ;
    \draw    (459.95,58.1) -- (516.45,58.57) ;
    \draw [shift={(519.45,58.6)}, rotate = 180.48] [fill={rgb, 255:red, 0; green, 0; blue, 0 }  ][line width=0.08]  [draw opacity=0] (10.72,-5.15) -- (0,0) -- (10.72,5.15) -- cycle    ;
    \draw [color={rgb, 255:red, 128; green, 128; blue, 128 }  ,draw opacity=1 ]   (230.47,58.1) -- (230.67,77.6) ;
    \draw [shift={(230.7,80.6)}, rotate = 269.43] [fill={rgb, 255:red, 128; green, 128; blue, 128 }  ,fill opacity=1 ][line width=0.08]  [draw opacity=0] (8.93,-4.29) -- (0,0) -- (8.93,4.29) -- cycle    ;
    \draw    (459.95,58.1) -- (550.5,58.53) ;
    \draw [color={rgb, 255:red, 128; green, 128; blue, 128 }  ,draw opacity=1 ]   (188.9,58.1) -- (239.88,57.8) ;
    \draw [color={rgb, 255:red, 128; green, 128; blue, 128 }  ,draw opacity=1 ]   (408.97,58.4) -- (459.95,58.1) ;
    \draw [color={rgb, 255:red, 128; green, 128; blue, 128 }  ,draw opacity=1 ] [dash pattern={on 0.84pt off 2.51pt}]  (239.88,57.8) -- (280.2,58.1) ;
    \draw [color={rgb, 255:red, 128; green, 128; blue, 128 }  ,draw opacity=1 ] [dash pattern={on 0.84pt off 2.51pt}]  (368.65,58.1) -- (408.97,58.4) ;
    \draw [color={rgb, 255:red, 128; green, 128; blue, 128 }  ,draw opacity=1 ]   (419.97,58.1) -- (420.17,77.6) ;
    \draw [shift={(420.2,80.6)}, rotate = 269.43] [fill={rgb, 255:red, 128; green, 128; blue, 128 }  ,fill opacity=1 ][line width=0.08]  [draw opacity=0] (8.93,-4.29) -- (0,0) -- (8.93,4.29) -- cycle    ;
    \draw   (189.75,109.38) .. controls (189.76,114.04) and (192.1,116.37) .. (196.77,116.36) -- (309.68,116.1) .. controls (316.35,116.09) and (319.69,118.41) .. (319.7,123.08) .. controls (319.69,118.41) and (323.01,116.07) .. (329.68,116.06)(326.68,116.06) -- (453.72,115.78) .. controls (458.39,115.77) and (460.71,113.43) .. (460.7,108.76) ;
    
    \draw (87.32,21.3) node [anchor=north west][inner sep=0.75pt]  [font=\large]  {$\underline{v}_{k}$};
    \draw (551.55,22.8) node [anchor=north west][inner sep=0.75pt]  [font=\large]  {$\underline{v}_{l}$};
    \draw (141.82,67.8) node [anchor=north west][inner sep=0.75pt]  [font=\large]  {$\underline{i}_{\mathrm{in}}$};
    \draw (502.8,67.3) node [anchor=north west][inner sep=0.75pt]  [font=\large]  {$\underline{i}_{\mathrm{out}}$};
    \draw (225.75,80.4) node [anchor=north west][inner sep=0.75pt]  [font=\small,color={rgb, 255:red, 128; green, 128; blue, 128 }  ,opacity=1 ]  {$\tilde{i}_{1}$};
    \draw (411.75,78.9) node [anchor=north west][inner sep=0.75pt]  [font=\small,color={rgb, 255:red, 128; green, 128; blue, 128 }  ,opacity=1 ]  {$\tilde{i}_{|\mathcal{O} |}$};
    \draw (215.75,24.5) node [anchor=north west][inner sep=0.75pt]  [font=\large,color={rgb, 255:red, 0; green, 0; blue, 0 }  ,opacity=0.8 ] [align=left] {{\fontfamily{ptm}\selectfont unmeasured subsystem}};
    \draw (255.25,122.78) node [anchor=north west][inner sep=0.75pt]    {$\tilde{i} =\sum \tilde{i}_{c} \ \ \forall \ c\ \in \ \mathcal{O}$};
    
    \end{tikzpicture}
        \caption{Grey-box subsystem of measured inputs and outputs with unmeasured consumers.}
        \label{fig:subproblem}
    \end{figure}

    The algorithm depicted in Fig. \ref{fig:flowchart} is employed to identify the switch status, phases, and parameters of a given \gls{LV} grid. The inputs required for this procedure are the taken measurements $\mathcal{M}$, the known configuration of the grid, i.e., the topology of graph $\mathcal{G}$ and the believed status of the switches, along with the types and typical parameters of cables that are used by the distribution system operator. The first step involves determining if the switches are open or closed,  and identifying the phases of the system. Once the phases are assigned and the measurement data is transformed into symmetrical components using \eqref{eq:symmcompiv}, one of two methods described in detail below can be employed. The first involves estimating the total impedance of the system, followed by the assignment of cable types to individual segments, and the second method is based on an optimization model. Upon completion of the procedure, the algorithm returns the estimated switch positions, phase assignments for the measurements, and the total impedance of the system, along with the impedance or cable types used in individual cable segments.
    
    \tikzstyle{io} = [trapezium, trapezium left angle=70, trapezium right angle=110, minimum width=3cm, minimum height=1cm, text centered, draw=black, fill=white!10, text width=5cm]
    \tikzstyle{process} = [rectangle, minimum width=3cm, minimum height=1cm, text centered, draw=black, fill=white!10, text width=6cm]
    \tikzstyle{decision} = [diamond, minimum width=3cm, minimum height=1cm, text centered, draw=black, fill=green!30]
    \tikzstyle{arrow} = [thick,->,>=stealth]
    
    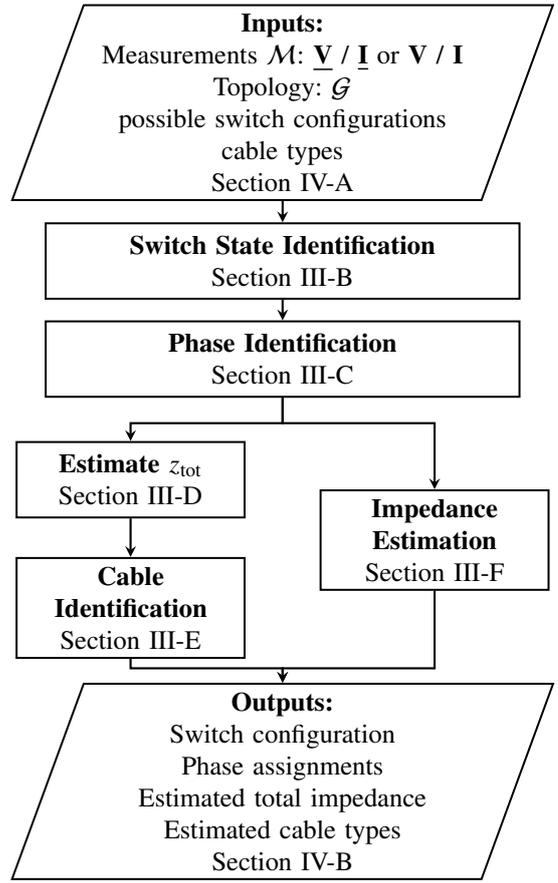
\begin{figure}
        \centering
        \begin{tikzpicture}[node distance=0.8cm]
            \node (in1) [io] {\textbf{Inputs:} \\ Measurements $\mathcal{M}$: $\underline{\mathbf{V}}$ /  $\underline{\mathbf{I}}$ or $\mathbf{V}$ / $\mathbf{I}$ \\
            Topology: $\mathcal{G}$ \\
            possible switch configurations \\
            cable types \\
            Section \ref{sec:testsystemmeasurements}};
            \node (pro1) [process, below of=in1, yshift=-1.3cm] {\textbf{Switch State Identification} \\
            Section \ref{sec:switchstateidentification}};
            \node (pro2) [process, below of=pro1, yshift=-.5cm]
            {\textbf{Phase Identification}\\ Section \ref{sec:phaseidentification}};
            \node (pro3) [process, below of=pro2, yshift=-0.8cm, text width=2.0cm, xshift=-2cm]{\textbf{Estimate $z_{\mathrm{tot}}$} \\ Section \ref{sec:totalimpedanceidentification}};
            \node (pro4) [process, below of=pro3, yshift=-0.9cm, text width=2.0cm]{\textbf{Cable \\ Identification} \\ Section \ref{sec:cabletypeidentification}};
            \node (pro5) [process, below of=pro2, yshift=-1.6cm, text width=2.0cm, xshift=+2cm]{\textbf{Impedance \\ Estimation} \\ Section \ref{sec:closedformimpedanceestimation}};
            \node (out1) [io, below of=pro4, yshift=-1.5cm, xshift=+2cm] {\textbf{Outputs:} \\ Switch configuration \\ Phase assignments \\ Estimated total impedance \\ Estimated cable types \\ Section \ref{sec:results}};
            \draw [arrow] (in1) -- (pro1);
            \draw [arrow] (pro1) -- (pro2);
            \draw [arrow] (pro2) -- (0,-4.25) -| (pro3);
            \draw [arrow] (pro3) -- (pro4);
            \draw [arrow] (pro4) |- (0,-7.5) -- (out1);
            \draw [arrow] (pro2) -- (0,-4.25) -| (pro5);
            \draw [arrow] (pro5) |- (0,-7.5) -- (out1);
        \end{tikzpicture}
        \caption{Flow chart of identification procedure.}
        \label{fig:flowchart}
    \end{figure}

    \subsection{Switch State Identification}
    \label{sec:switchstateidentification}
    Given the available measurements and potential switch configurations, the switch states, i.e., open or closed, can be determined by comparing the current measurements with threshold values. It is assumed that a switch is present at each grid junction, which can be utilized to alter the connection in the junction from open loop to short circuit. Since the incoming and outgoing currents of all phases are measured, the switch position can be inferred by comparing the incoming and outgoing currents as follows:

    \begin{equation}
         \sum_k^{\{ \text{A}, \text{B}, \text{C} \}} \max(|\underline{i}_{\mathrm{in, n, k}}|, |\underline{i}_{\mathrm{out, n, k}}|) \geq \epsilon \qquad \forall n \in \mathcal{M} 
        \label{eq:switchstatus1}
    \end{equation}
    where $\mathcal{M}$ denotes the set of measured nodes. Thus, if this holds, the switch is considered closed. The parameter $\epsilon$ has to be chosen sufficiently large to prevent faulty detections from measurement noise. In case the load at all downstream nodes is zero, an incorrect detection of the open state may occur. To prevent this, a threshold based on the summation of several measurements over time is utilized as follows:
    \begin{equation}
         \sum_{t=1}^T\sum_k^{\{ \text{A}, \text{B}, \text{C} \}} \max(|\underline{i}_{\mathrm{in, n, k}}(t)|, |\underline{i}_{\mathrm{out, n, k}}(t)|) \geq \epsilon \qquad \forall n \in \mathcal{M} 
        \label{eq:switchstatus2}
    \end{equation}
    
    \subsection{Phase Identification}
    \label{sec:phaseidentification}
    \gls{PHI} refers to the accurate allocation of measurements of currents and voltages for specific system phases to their estimated phases, which plays a critical role in subsequent algorithms for \gls{PI}. Incorrect phase assignments may result in the misinterpretation of measurements for different system phases, leading to inaccuracies in the \gls{PI} methods. \gls{PHI} involves distinguishing between two types of typical measuring devices used in distribution grids. When \glspl{PMU} are available, phasors can be reconstructed for all measurements, enabling the application of methods that consider the angles of current and voltage measurements. However, the more common meters used in distribution grids are Non-Phasor Measurement Units (\glspl{NPMU}), which only measure the magnitudes of the respective voltage and currents, as well as the phase offset of the measurements of the same meter. As a result, the current and voltage angles cannot be used to assign phases, since the relationship to the reference node angle is unknown. Consequently, two different approaches are applied depending on the available measurement devices, as explained next.
    
    \subsubsection{PHI from PMUs}
    \label{subsubsec:phaseidentificationphasors}
    If measurements from \glspl{PMU} are available, the system phase can be assigned by clustering the measured voltage angles. Since voltage angle deviations in distribution grids are typically small, i.e., only a few degrees, three clusters are formed around 0°, 120°, and -120°, where the angle of 0° pertains to the voltage angle measured at the reference node, even though there might be significant current unbalances. Thus, even though there might be significant current inbalances, it can still be assumed that the voltages are balanced. Thus, the following method is suggested for \gls{PHI} in this case:
    \begin{enumerate}[I.]
        \item Define the cluster centers $\mu_{\text{A}} = 0.0 ^{\circ}$, $\mu_{\text{B}} = -120.0 ^{\circ}$, and $\mu_{\text{C}} = 120.0 ^{\circ}$.
        \item Set the phase of each measurement to the phase $k \in \{\text{A}, \text{B}, \text{C}\}$ with least absolute distance $|\mu_k - \overline{\theta}|$, where $\overline{\theta}$ is the average voltage angle for a given cluster.
    \end{enumerate}
    If significant measurement errors or outliers are anticipated in the angle measurements, clustering can be conducted using the measurements from several time steps.
    
    \subsubsection{PHI from NPMUs}
    \label{subsubsec:phaseidentificationnonphasors}
    In this case, the voltage and current measurements only contain magnitudes. As a result, the phase angle between nodes is unknown, which prevents the clustering of measurements. However, the correlation between the voltage and current time series of the two nodes can be used to identify the phase. Thus, since the two nodes are electrically connected, it can be assumed that voltages and currents of the same phase show a higher correlation compared to nodes of different phases. The resulting correlation matrix is a square matrix containing the cross-correlation between the measurements at node $k$ with the measurements at node $l$.

    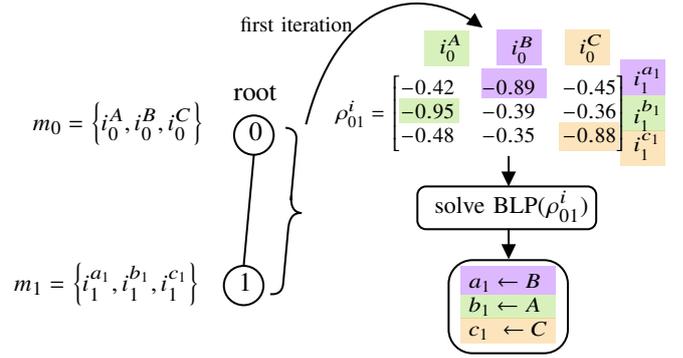
\begin{figure}   
        \begin{tikzpicture}[x=0.75pt,y=0.75pt,yscale=-0.85,xscale=0.85]
        
        \draw   (137,81.63) .. controls (137,75.2) and (142.2,70) .. (148.63,70) .. controls (155.05,70) and (160.25,75.2) .. (160.25,81.63) .. controls (160.25,88.05) and (155.05,93.25) .. (148.63,93.25) .. controls (142.2,93.25) and (137,88.05) .. (137,81.63) -- cycle ;
        
        \draw   (131,170.63) .. controls (131,164.2) and (136.2,159) .. (142.63,159) .. controls (149.05,159) and (154.25,164.2) .. (154.25,170.63) .. controls (154.25,177.05) and (149.05,182.25) .. (142.63,182.25) .. controls (136.2,182.25) and (131,177.05) .. (131,170.63) -- cycle ;
        \draw    (142.63,159) -- (148.63,93.25) ;
        \draw   (158.23,175.7) .. controls (162.88,176.13) and (165.41,174.02) .. (165.84,169.37) -- (168.79,137.08) .. controls (169.4,130.44) and (172.02,127.33) .. (176.67,127.76) .. controls (172.02,127.33) and (170,123.8) .. (170.61,117.16)(170.34,120.15) -- (173.56,84.87) .. controls (173.99,80.22) and (171.88,77.69) .. (167.23,77.27) ;
        \draw    (179.23,75.27) .. controls (196.6,5.51) and (241.12,-11.25) .. (280.54,16.34) ;
        \draw [shift={(282.95,18.08)}, rotate = 216.87] [fill={rgb, 255:red, 0; green, 0; blue, 0 }  ][line width=0.08]  [draw opacity=0] (8.93,-4.29) -- (0,0) -- (8.93,4.29) -- cycle    ;
        \draw  [draw opacity=0][fill={rgb, 255:red, 144; green, 19; blue, 254 }  ,fill opacity=0.3 ][dash pattern={on 0.84pt off 2.51pt}] (282.23,42.3) -- (320.47,42.3) -- (320.47,60) -- (282.23,60) -- cycle ;
        \draw  [draw opacity=0][fill={rgb, 255:red, 126; green, 211; blue, 33 }  ,fill opacity=0.3 ][dash pattern={on 0.84pt off 2.51pt}] (234,60) -- (269.23,60) -- (269.23,74.7) -- (234,74.7) -- cycle ;
        \draw  [draw opacity=0][fill={rgb, 255:red, 144; green, 19; blue, 254 }  ,fill opacity=0.3 ][dash pattern={on 0.84pt off 2.51pt}] (291,19.1) -- (317.23,19.1) -- (317.23,40.1) -- (291,40.1) -- cycle ;
        \draw  [draw opacity=0][fill={rgb, 255:red, 245; green, 166; blue, 35 }  ,fill opacity=0.3 ][dash pattern={on 0.84pt off 2.51pt}] (328,74) -- (363.23,74) -- (363.23,91) -- (328,91) -- cycle ;
        \draw   (244.65,117.06) .. controls (244.65,114.27) and (246.92,112) .. (249.71,112) -- (344.59,112) .. controls (347.38,112) and (349.65,114.27) .. (349.65,117.06) -- (349.65,132.24) .. controls (349.65,135.03) and (347.38,137.3) .. (344.59,137.3) -- (249.71,137.3) .. controls (246.92,137.3) and (244.65,135.03) .. (244.65,132.24) -- cycle ;
        \draw  [draw opacity=0][fill={rgb, 255:red, 144; green, 19; blue, 254 }  ,fill opacity=0.3 ][dash pattern={on 0.84pt off 2.51pt}] (364,37.1) -- (390.23,37.1) -- (390.23,58.1) -- (364,58.1) -- cycle ;
        \draw  [draw opacity=0][fill={rgb, 255:red, 126; green, 211; blue, 33 }  ,fill opacity=0.3 ][dash pattern={on 0.84pt off 2.51pt}] (249,20.1) -- (275.23,20.1) -- (275.23,41.1) -- (249,41.1) -- cycle ;
        \draw  [draw opacity=0][fill={rgb, 255:red, 245; green, 166; blue, 35 }  ,fill opacity=0.3 ][dash pattern={on 0.84pt off 2.51pt}] (364,79.1) -- (390.23,79.1) -- (390.23,100.1) -- (364,100.1) -- cycle ;
        \draw  [draw opacity=0][fill={rgb, 255:red, 245; green, 166; blue, 35 }  ,fill opacity=0.3 ][dash pattern={on 0.84pt off 2.51pt}] (331.23,19) -- (357.47,19) -- (357.47,40) -- (331.23,40) -- cycle ;
        \draw  [draw opacity=0][fill={rgb, 255:red, 126; green, 211; blue, 33 }  ,fill opacity=0.3 ][dash pattern={on 0.84pt off 2.51pt}] (364,58.1) -- (390.23,58.1) -- (390.23,79.1) -- (364,79.1) -- cycle ;
        \draw    (298.23,93.98) -- (298.23,109.1) ;
        \draw [shift={(298.23,112.1)}, rotate = 270] [fill={rgb, 255:red, 0; green, 0; blue, 0 }  ][line width=0.08]  [draw opacity=0] (8.93,-4.29) -- (0,0) -- (8.93,4.29) -- cycle    ;
        \draw   (263,169.6) .. controls (263,161.87) and (269.27,155.6) .. (277,155.6) -- (319,155.6) .. controls (326.73,155.6) and (333,161.87) .. (333,169.6) -- (333,196.7) .. controls (333,204.43) and (326.73,210.7) .. (319,210.7) -- (277,210.7) .. controls (269.27,210.7) and (263,204.43) .. (263,196.7) -- cycle ;
        \draw  [draw opacity=0][fill={rgb, 255:red, 144; green, 19; blue, 254 }  ,fill opacity=0.3 ][dash pattern={on 0.84pt off 2.51pt}] (270,159.1) -- (325.83,159.1) -- (325.83,176.1) -- (270,176.1) -- cycle ;
        \draw  [draw opacity=0][fill={rgb, 255:red, 126; green, 211; blue, 33 }  ,fill opacity=0.3 ][dash pattern={on 0.84pt off 2.51pt}] (270,176.1) -- (325.83,176.1) -- (325.83,190.1) -- (270,190.1) -- cycle ;
        \draw  [draw opacity=0][fill={rgb, 255:red, 245; green, 166; blue, 35 }  ,fill opacity=0.3 ][dash pattern={on 0.84pt off 2.51pt}] (270,190.1) -- (325.83,190.1) -- (325.83,205.1) -- (270,205.1) -- cycle ;
        \draw    (298.23,136.98) -- (298.23,152.1) ;
        \draw [shift={(298.23,155.1)}, rotate = 270] [fill={rgb, 255:red, 0; green, 0; blue, 0 }  ][line width=0.08]  [draw opacity=0] (8.93,-4.29) -- (0,0) -- (8.93,4.29) -- cycle    ;
        
        \draw (138,161) node [anchor=north west][inner sep=0.75pt]   [align=left] {{\fontfamily{ptm}\selectfont 1}};
        \draw (17,60.4) node [anchor=north west][inner sep=0.75pt]  [font=\small]  {${\textstyle m_{0} =\left\{i_{0}^{A} ,i_{0}^{B} ,i_{0}^{C}\right\}}$};
        \draw (6,155.4) node [anchor=north west][inner sep=0.75pt]  [font=\small]  {${\textstyle m_{1} =\left\{i_{1}^{a_{1}} ,i_{1}^{b_{1}} ,i_{1}^{c_{1}}\right\}}$};
        \draw (195,44.4) node [anchor=north west][inner sep=0.75pt]  [font=\footnotesize]  {$\rho _{01}^{i} =\begin{bmatrix}
        -0.42 & -0.89 & -0.45\\
        -0.95 & -0.39 & -0.36\\
        -0.48 & -0.35 & -0.88
        \end{bmatrix}$};
        \draw (256,20.4) node [anchor=north west][inner sep=0.75pt]  [font=\footnotesize]  {$i_{0}^{A}$};
        \draw (298,21.4) node [anchor=north west][inner sep=0.75pt]  [font=\footnotesize]  {$i_{0}^{B}$};
        \draw (338,20.4) node [anchor=north west][inner sep=0.75pt]  [font=\footnotesize]  {$i_{0}^{C}$};
        \draw (370,39.4) node [anchor=north west][inner sep=0.75pt]  [font=\footnotesize]  {$i_{1}^{a_{1}}$};
        \draw (370,59.4) node [anchor=north west][inner sep=0.75pt]  [font=\footnotesize]  {$i_{1}^{b_{1}}$};
        \draw (370,78.4) node [anchor=north west][inner sep=0.75pt]  [font=\footnotesize]  {$i_{1}^{c_{1}}$};
        \draw (139,10) node [anchor=north west][inner sep=0.75pt]   [align=left] {{\footnotesize {\fontfamily{ptm}\selectfont first iteration}}};
        \draw (265,160.5) node [anchor=north west][inner sep=0.75pt]  [font=\footnotesize]  {$ \begin{array}{l}
        a_{1}\leftarrow B\\
        b_{1}\leftarrow A\\
        c_{1} \ \leftarrow C
        \end{array}$};
        \draw (135,50) node [anchor=north west][inner sep=0.75pt]   [align=left] {{\fontfamily{ptm}\selectfont root}};
        \draw (144,72) node [anchor=north west][inner sep=0.75pt]   [align=left] {{\fontfamily{ptm}\selectfont 0}};
        \draw (253,113) node [anchor=north west][inner sep=0.75pt]  [font=\small] [align=left] {{\fontfamily{ptm}\selectfont solve \gls{BLP}}($\displaystyle \rho _{01}^{i}$)};
        \end{tikzpicture}
        
        \caption{Example for PHI with NPMUs.}
        \label{fig:examplephaseassignmentNPMU}
    \end{figure}

    \begin{algorithm}
        \begin{algorithmic}
            \State \textbf{Inputs:} $\text{ID}_{\text{root}}$, $\mathcal{G}$, phases root, measurements, leaf nodes $\mathcal{L}$
            \State $\mathcal{P} \leftarrow \{\text{ID}_{\text{root}} \}$
            \While{ $\mathcal{P} \neq \varnothing$}
            \ForAll{$p \in \mathcal{P}$}
            \State $\mathcal{C} \leftarrow \mathrm{children}(p, \mathcal{G})$
            \ForAll{$c \in \mathcal{C}$} 
            \If{$c \notin \mathcal{L}$}
            \State estimate phases using $\rho_{p,c}^{i}$ solving \eqref{eq:blp}
            \Else{}
            \State estimate phases using $-\rho_{p,c}^{v}$ solving \eqref{eq:blp} 
            \EndIf
            \EndFor
            \EndFor
            \State $\mathcal{P} \leftarrow \mathcal{C} - (\mathcal{C} \cap \mathcal{L})$
            \EndWhile
        \end{algorithmic}
        \caption{\gls{NPMU} \gls{PHI}}
        \label{algo:nonpmuphase}
    \end{algorithm}
    Algorithm \ref{algo:nonpmuphase} shows the pseudo-code to estimate the phase assignments of all measurements in the system. It assumes that the phase assignment is given at the root node, which is the connection to the \gls{MV} grid. It then traverses the graph of the tree spanned by the measurements until the leaf nodes are reached. For each parent-child node pair, the following optimization problem is solved:

    \begin{equation}
        \begin{aligned}
            \underset{x}{\text{min}} \quad & \qquad \mathrm{vec}(\rho_{k,l}^i)^\intercal \cdot \mathbf{x}\\
            \begin{matrix} \text{s.t.} \\ \\ \\ \\ \end{matrix} \quad & \begin{bmatrix} 
            \mathbb{1}_{1\times 3} & & \\ 
            & \mathbb{1}_{1\times 3} & \\ 
            & & \mathbb{1}_{1\times 3} \\
            \mathbb{1}_3 & \mathbb{1}_3 & \mathbb{1}_3 \\
            \end{bmatrix}_{6 \times 9} \cdot \mathbf{x} = \mathbb{1}_{6 \times 1} \\ 
        \end{aligned}
    \label{eq:blp}
    \end{equation}
    where $\mathbf{x} = \left[ x_1, \dots, x_9 \right]^\intercal$, $x_i \in \{0, 1\}$, and $\mathrm{vec}(\rho_{k,l}^i) \in \mathbb{R}^{1\times 9}$ is the column-wise vectorized correlation matrix of current measurements between parent and child node and aims at minimizing the summed correlation over all system phases by solving this Binary Linear Programming (\gls{BLP}) problem. The solution to this optimization problem is a binary assignment of the known, or previously estimated, phase assignments of the parent node to the unknown phase assignments of the child node. The equivalent problem can be formulated for the voltage correlations by multiplying the objective function with minus one, thus turning the problem into a maximization. Due to the higher robustness against noise on the measurements, the correlation on current measurements is chosen for this procedure, since voltage magnitudes at different nodes are all close to one per unit, while currents magnitudes cover a larger range and can show unbalances, making them more robust against measurement noise. Concluding that the current flowing into the leaf nodes is zero when no loads are attached to the leafs, the voltage correlation is used for phase assignment at the these nodes. 
    
    Fig. \ref{fig:examplephaseassignmentNPMU} shows an example of the application of the proposed \gls{PHI} procedure for \glspl{NPMU} on a graph with two measured nodes. Thus, the correlation of current measurements at Nodes 0 and 1 are calculated based on solving the \gls{BLP} \eqref{eq:blp}, with the estimated phases being assigned to the variables $a_1$, $b_1$ and $c_1$. For the next nodes, the same procedure is applied sequentially until the last node of the branch is reached.
    
    \subsection{Total Impedance Estimation}
    \label{sec:totalimpedanceidentification}
    The estimation of the total impedance aims at estimating the impedance of the subsystems, as shown in Fig. \ref{fig:subproblem}. The regression method presented in \cite{dejongh2023parameter} is used for this estimation, based on the following assumptions:
    \begin{enumerate}
        \item For each observable subsystem with boundary nodes $k$ and $l$, the voltage decreases along the branch $|\underline{V}_k| \geq |\underline{V}_l|$ and the inflowing current has a higher magnitude than the outflowing current $|\underline{i}_{\mathrm{in}}| \geq |\underline{i}_{\mathrm{out}}|$.
        \item For all consumers $|\underline{i}_k| \geq 0$, $\forall k \in \mathcal{O}$.
    \end{enumerate}
    The first assumption can be ensured by switching nodes $k$ and $l$ in case the assumption does not hold. The second assumption considers that there is no grid current injection at the consumer site, i.e., no generation.

    The participation factor is defined as the proportion of the out-flowing current $\underline{i}_{\mathrm{out}}$ to the current $\underline{i}_{\mathrm{in}}$ flowing into the subsystem:
    \begin{equation}
        f = \frac{|\underline{i}_{\mathrm{out}}|}{|\underline{i}_{\mathrm{in}}|}
    \end{equation}
    which is in the range $0 \leq f \leq 1$. This allows the formulation of upper and lower bounds on the subsystem impedances. Under the assumptions made, this allows the formulation of bounds, which can be determined by means of limit value considerations, as follows:
    \begin{equation}
        \underline{z}_{kl, \mathrm{lb}} = \frac{\underline{v}_{k} - \underline{v}_{l}}{\underline{i}_{\mathrm{in}}}
        \label{eq:zlb}
    \end{equation}
    
    \begin{equation}
        \underline{z}_{kl, \mathrm{ub}} = \frac{\underline{v}_{k} - \underline{v}_{l}}{\underline{i}_{\mathrm{out}}}
        \label{eq:zub}
    \end{equation}
    where $\underline{z}_{kl, \mathrm{lb}}$ and $\underline{z}_{kl, \mathrm{ub}}$ are the respective lower and upper bounds between which the true subsystem impedance lies. It is evident that for a participation factor of $f=1$, \eqref{eq:zlb} and \eqref{eq:zub} are identical such that the actual total subsystem impedance is obtained. However, this case may not occur in reality, so that there is an interval of possible total impedances. As in \cite{dejongh2023parameter}, a regression problem that extrapolates the total impedance of the bounds in \eqref{eq:zlb} and provides an estimate of the total impedance from measured data can be formulated.
    
    Further adjustments must be made to apply the regression method to the multiphase system at hand. Thus, first the methodology is applied only to the positive sequence component, marked by superscript ones, where the measurements are converted to symmetric components using \eqref{eq:symmcompiv}. In addition, only the magnitude of the respective voltages and currents are considered for the consideration of \gls{NPMU} measurements. The resulting optimization problem can then be defined as follows:

    \begin{equation}
        \underset{\beta}{\min} \sum_{t=1}^{T} w(t) \cdot \left( z_{\mathrm{lb}}^{1}(t) - f^1(t) \cdot \beta_0 - \beta_1 \right)^2
        \label{eq:wlsobjective}
    \end{equation}
    where $T$ is the number of measured time steps and $\beta_0$ and $\beta_1$ are the slope and intercept of the regression line, respectively. Additional weighting factors $w(t)$ can be added to give more importance to more informative measurements with higher participation factors. The Weighted Least Squares (\gls{WLS}) problem \eqref{eq:wlsobjective} is solved and leads to the following estimate of the total impedance:

    \begin{equation}        \hat{z}_{\mathrm{tot}}^{\mathrm{1}} = \beta_0 + \beta_1
        \label{eq:ztotest}
    \end{equation}
    which is obtained from the regression line evaluated at $f=1$ and gives an estimate of the magnitude of the total positive sequence impedance $\hat{z}_{\mathrm{tot}}^{\mathrm{1}}$ of the respective subsystem.

    \subsection{Cable Type Identification}
    \label{sec:cabletypeidentification}
    The goal of cable type identification is to use the previously estimated total impedance of a subsystem to estimate the cable types used in each segment in that subsystem. The following three assumptions are made for estimating the cable types:
    
    \begin{enumerate}
        \item The geographical cable lengths or estimates thereof are known from, for example, \gls{GIS}, and their lengths are denoted as $\ell$.
        \item The estimate of the total subsystem impedance of the positive sequence is available, as described in Section \ref{sec:totalimpedanceidentification}.
        \item The candidate cable types for the given distribution system are known.
    \end{enumerate}
    Thus, the estimates of cable types can then be calculated by solving the following optimization problem:

    \begin{align}
        \underset{z^{1'}_k}{\min} \quad & \left\| \hat{z}^1_\mathrm{tot} - \sum_{k=1}^{\mathcal{S}} z^{1'}_k\cdot \ell_k \right\|^2 \label{eq:objeasy} \\ 
        \text{s.t.} \quad & z^{1'}_k \in \{z^{1'}_{c_1}, z^{1'}_{c_2}, \dots \} \label{eq:setmembership} \\
        & z^{1'}_k \leq z^{1'}_{k+1} \leq \dots \leq z^{1'}_{|\mathcal{S}|} \quad \forall k \in \mathcal{S} \label{eq:inequalityeasy}
    \end{align}
    which is a Mixed-Integer Quadratic Programming (\gls{MIQP}) problem with categorical per unit length variables $z^{1'}_k$ and linear equality and inequality constraints, where $\mathcal{S}$ denotes the set of all cable segments in the respective branch. The objective function aims at minimizing the squared residual of estimated total impedance and the combination of chosen cable types $c_1, c_2, \dots$, multiplied by the assumed lengths of lines $\ell_k$. The constraint \eqref{eq:setmembership} ensures that only available cable type parameters $z^{1'}_{c_i}$ are chosen. The inequality constraint \eqref{eq:inequalityeasy} restricts the solution to increasing resistances per length along the branch.
    
    \subsection{Closed-Form Impedance Estimation}
    \label{sec:closedformimpedanceestimation}
    As depicted in Fig. \ref{fig:flowchart}, it is possible to get an estimation of the total impedance and the types of cables used through a different optimization problem. This approach assumes that the number of cable segments, available cable types, and lengths of the segments are known. The primary objective is to determine an estimate of the impedance of all the segments present in the subsystem under consideration, using measured voltages and currents at the input and output of the respective subsystem for all $T$ time steps as input data for this purpose. Thus, let $n_s=|\mathcal{S}|$ be the number of segments and $\mathcal{C}$ a set that contains all available impedances per length of the positive sequence. Then, the following optimization problem can be posed:

    \begin{align}
        \underset{z^{1}_{k}, i^{1}_{k}}{\min} \quad & \sum_{t=1}^{T} \bigg( \Delta v^{1}(t) + \sum_{k=1}^{n_s}i^{1}_{k}(t) \cdot z^{1}_{k}\bigg)^2 \label{eq:objectiveGEN} \\
        \text{s.t.:} \quad & i^{1}_{1}(t) \geq i^{1}_{2}(t) \geq \dots \geq i^{1}_{n_s}(t) \qquad \quad \;\;\; \forall t \in T \label{eq:constraintGENdecliningflow} \\
         & \Delta t \sum_{t=1}^{T} \hat{v}^{1}_k(t) \cdot \big(i^{1}_{k+1}(t) - i^{1}_{k}(t)\big) = E_k \;\; \forall k \in \mathcal{S}\char`\\ n_s\label{eq:constraintGENesum} \\
         & z^{1}_{k} = z^{1'}_{k} \cdot \ell_k \qquad \qquad \qquad z^{1'}_{k}\in \mathcal{C} \; \forall k \in \mathcal{S} \label{eq:constraintGENrsrcomb} \\
    \end{align}
    \text{where:}
    \begin{align}
        &\Delta v^{1}(t) = -v^{1}_{\mathrm{in}}(t) + v^{1}_{\mathrm{out}}(t)  \qquad \quad \; \forall t \in T \label{eq:dVGEN1} \\
        & i^{1}_{0,1}(t) = i^{1}_{\mathrm{in}}(t) \qquad \qquad \qquad \qquad \forall t \in T \label{eq:dVGEN2}\\
        & i^{1}_{n_s}(t) = i^{1}_{\mathrm{out}}(t) \qquad \qquad \qquad \qquad \forall t \in T \label{eq:dVGEN3}\\
        & z^{1}_{k} \geq 0 \qquad \qquad \qquad \qquad \qquad \quad \; \forall k \in \mathcal{S} \label{eq:dVGEN4}        
    \end{align}
    This optimization problem aims to minimize the squared deviation between the measured positive sequence voltage drop $\Delta v^{1}(t)$ and the estimated voltage drop in the subsystem for all time steps $T$. The estimated voltage drop is determined by the product of the impedance and the currents flowing through each segment of the system, which defines the objective function \eqref{eq:objectiveGEN}. In order to ensure the validity of the problem, it is necessary to add constraints that limit the values of the impedances and prevent unrealistic assignments of currents. Furthermore, since no generation is considered, the current flowing through the system decreases over each branch as per \eqref{eq:constraintGENdecliningflow}. Assuming that the energy consumption $E_k$ is available from energy measurements, and the estimated voltage at each node is given, \eqref{eq:constraintGENesum} can be defined. The values of the impedances are restricted to a set of known cables $\mathcal{C}$ as per \eqref{eq:constraintGENrsrcomb}. Thus, this is an \gls{MIQP} optimization problem.

    \section{Simulation and Results}
    \label{sec:simandresults}
    \subsection{Test system and Measurements}
    \label{sec:testsystemmeasurements}
\begin{figure}
        \centering
        \includegraphics[width=0.8\columnwidth]{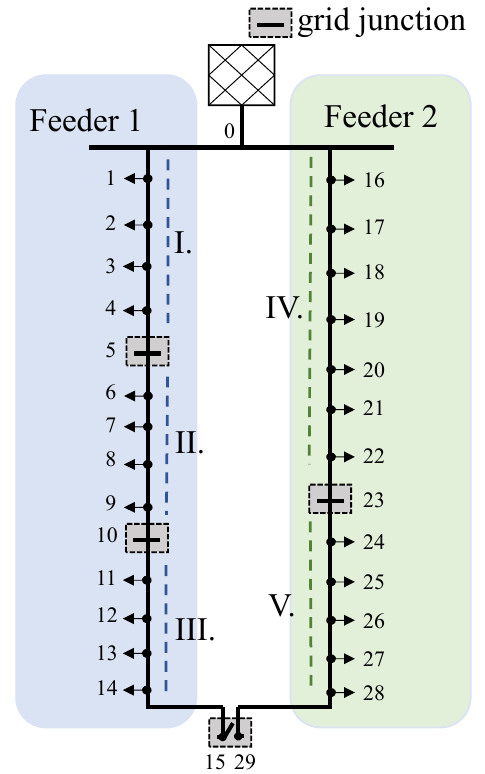}
        \caption{Test grid for SI study.}
        \label{fig:benchobservability}
    \end{figure}

    To compare the proposed \gls{SI} methods for \gls{LV} grids with limited measurements, the realistic test grid shown in Fig. \ref{fig:benchobservability} is used. The grid is based on a real \gls{LV} grid in Germany supplying a residential area. It is a TNC-S system with a nominal phase-to-phase voltage of 400 V, and is supplied by a 20 kV \gls{MV} grid. The grid consists of 30 nodes, 24 of which are feeders to individual households. It is assumed that all households are connected through all three system phases. As shown in gray in Fig. \ref{fig:benchobservability}, there are four grid junctions at the grid at nodes 5, 10, 15, 29 and 23. The grid junctions at nodes 15 and 19/29 contain a switch such that the grid can be operated either as a radial grid or occasionally as a meshed grid. The cable parameters and configuration used can be seen in Table \ref{tab:subproblems}.

    \begin{table}[h!]
        \centering
        \caption{Subsystems in test \gls{LV} grid.}
        \begin{tabular}{|c|c|c|c|c|c|c|}
            \hline
             \textbf{Subsystem} & \textbf{k} & \textbf{l} & $r_{kl, \text{true}}$ & $x_{kl, \text{true}}$ & \textbf{Lengths} & \textbf{Linetypes} \\
             & & & [m$\Omega$] & [m$\Omega$] & [m] & [mm$^2$] \\
             \hline
             \hline
             I & 0 & 5 & 15.71 & 6.04 & 61.31 & 150 \\
             & & & & & 1.66 & 150 \\
             & & & & & 7.99 & 150 \\
             & & & & & 1.87 & 150 \\
             & & & & & 2.69 & 150 \\
             \hline
             II & 5 & 10 & 10.21 & 3.63 & 16.14 & 120\\
             & & & & & 1.61 & 120\\
             & & & & & 8.18 & 120\\
             & & & & & 1.57 & 120\\
             & & & & & 17.91 & 120\\
             \hline
             III & 10 & 15 & 39.32 & 5.08 & 5.39 & 50 \\
             & & & & & 1.16 & 50 \\
             & & & & & 19.94 & 50 \\
             & & & & & 4.80 & 50 \\
             & & & & & 29.96 & 50 \\
             \hline
             IV & 0 & 23 & 20.22 & 7.56 & 27.93 & 150 \\
             & & & & & 16.24 & 150\\
             & & & & & 5.49 & 150 \\
             & & & & & 5.84 & 150 \\
             & & & & & 5.45 & 150 \\
             & & & & & 8.78 & 120 \\
             & & & & & 1.16 & 120 \\
             & & & & & 23.57 & 120 \\
             \hline
             V & 23 & 29 & 87.38 & 11.30 & 3.85 & 50\\
             & & & & & 27.11 & 50 \\
             & & & & & 6.06 & 50 \\
             & & & & & 24.06 & 50 \\
             & & & & & 7.60 & 50 \\
             & & & & & 67.43 & 50 \\
             \hline
        \end{tabular}
        \label{tab:subproblems}
    \end{table}

    It is assumed that measurements can only be obtained at nodes 0, 5, 10, 15, 23, and 29, which belong to the grid operator. Overall, the factor of measured to unmeasured nodes is $6/30=0.2$. For each measured node in the grid it is assumed that the voltages and all inflowing and outflowing currents at all phases A, B, and C are measured. 
    
    Two types of measurements may be available:
    \begin{enumerate}
        \item Measurements by \gls{PMU} that allow to synchronize the measurements at all nodes \cite{uPMU}.
        \item Measurements by \gls{NPMU} that only measure the magnitude of voltages and currents.
    \end{enumerate}
    For the first case, the measurements of the different nodes are synchronized using Global Positioniong System (\gls{GPS}), allowing the phase angle between voltages at the different grid nodes to be determined. In practice, the measurement of grid quantities by means of \glspl{PMU} in distribution grids has already been proven through so-called $\mu$-\glspl{PMU} \cite{uPMU}. In the second case, if measurements are obtained from \gls{NPMU}, only the magnitudes of currents and voltages are available. Due to the low accuracy of synchronization via Network Time Protocol (\gls{NTP}) \cite{NTP}, the voltage angles between individual voltages cannot be determined. Hence, \eqref{eq:voltagedropline} cannot be applied in this case, using the approximation \eqref{eq:voltagedroplinesimplified} instead. The test system can be divided into subsystems using the methodology from Section \ref{sec:systemidentificationalgorithm}, as shown in Fig.~\ref{fig:subproblem}. This leads to subsystems I, II, III, IV, and V, which are considered individually.
    
    \subsection{Results}
    \label{sec:results}

    \subsubsection{Switch State Estimation}
    \label{sec:resultsswitchstatestimation}
    The estimation of switch states, as described in Section \ref{sec:switchstateidentification}, is applied to all grid junctions of the benchmark grid. The correct switch positions and thus the correct topology can be determined by the proposed method and under typical measurement noise in all cases without error. This allows to assume the topology for all further estimations as given and error-free.
    
    \subsubsection{Phase Identification}
    \label{sec:resultsphaseidentification}
\begin{figure}[ht]
        \centering
        \includegraphics[width=\columnwidth]{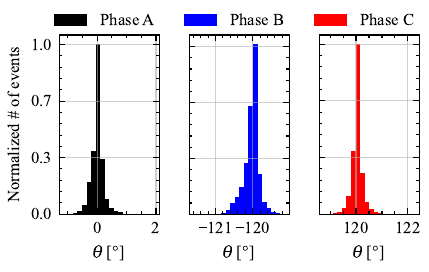}
        \caption{Histograms of all measured voltage angles for phases A, B and C.}
        \label{fig:histAngles}
    \end{figure}

    As discussed, the identification of the measurement phases depends on the type of measuring devices used. Consequently, two types of measurements must be distinguished for this purpose. Fig. \ref{fig:histAngles} shows the histograms for the case when measurements are obtained from \gls{PMU}. Observe that an unambiguous assignment of the phases can be carried out by means of clustering, even with strong measurement noise.
    
\begin{figure}[ht]
        \centering
        \includegraphics[width=\columnwidth]{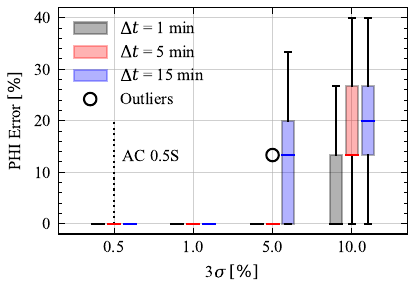}
        \caption{\gls{NPMU} \gls{PHI} error with varying $\Delta t$ and measurement noise $3\sigma$.}
        \label{fig:PIErrors}
    \end{figure}

    A more challenging problem is faced when measurements are obtained from \gls{NPMU}, leading to the application of Algorithm 1. Fig. \ref{fig:PIErrors} shows the effect of measurement noise on \gls{PHI} results for this case for different noise levels and data resolutions. For comparison, the typical accuracy class (\gls{AC}) for grid measurement devices 0.5S, which corresponds to $3\sigma \approx 0.5\%$, is plotted. Note that the methodology provides error-free predictions up to $3\sigma = 5\%$ for all $\Delta t$. Furthermore, observe that the robustness increases with lower sampling rate and the identification error only increases with higher noise.
    
    From the aforementioned results it can be concluded that, under realistic circumstances, both methods lead to error-free estimation of the system phases for the given scenario. Hence, a correct phase assignment can be assumed for all the methods discussed next.

    \subsubsection{Total Impedance Estimation}
    \label{sec:resultstotalresistanceestimation}
    The results of total impedance estimation of the subsystems I to IV using the methods presented in Section \ref{sec:totalimpedanceidentification} are evaluated here using the relative estimation error $100 \cdot (\hat{z}_{\text{tot}}-z_{\text{tot, true}})/z_{\text{tot, true}}$. Table \ref{tab:rtotesterrors} presents the estimation errors obtained for the tree topology, with the switch open between nodes 15 and 29, and for the meshed topology with a closed switch. Observe that the estimation errors for the impedances are similar for both configurations and increase for subsystems that are further away from the root node, due to the assumptions in Section \ref{sec:totalimpedanceidentification}. On the one hand, the disregard of the reactive power becomes apparent, as well as the linear approximation in \eqref{eq:ztotest}. Furthermore, note that the estimation error is always negative, indicating that the estimated value for the total impedance is lower than the actual total impedance for all cases considered, due to the linear approximation of the voltage drop.

    \begin{table}[ht]
        \centering
        \caption{Subsystem total impedance estimation errors}
        \begin{tabular}{|c|c|c|}
        \hline
        \textbf{Subsystem} & \textbf{Tree} & \textbf{Meshed} \\
        \hline
        I &  $-0.315\%$ & $-0.337\%$ \\
        \hline
        II &  $-5.492\%$ &  $-5.477\%$ \\
        \hline
        III &  - &  $-3.436\%$ \\
        \hline
        IV &  $-2.143\%$ & $-2.425\%$ \\
        \hline
        V &  - & $-14.782\%$ \\
        \hline
        \end{tabular}
        \label{tab:rtotesterrors}
    \end{table}

\begin{figure}[ht]
        \centering
        \includegraphics[width=\columnwidth]{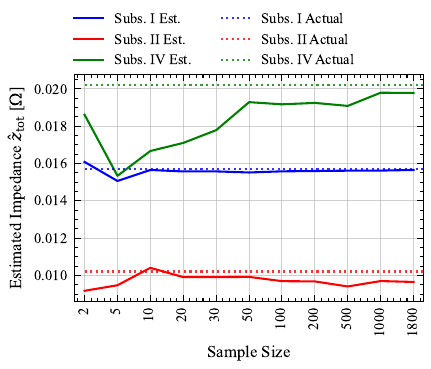}
        \caption{Influence of sample size on estimates of $\hat{z}_{\text{tot}}$ for $\Delta t = 1\; \mathrm{min}$.}
        \label{fig:ssizertot}
    \end{figure}

    In addition to the quality of the PHI estimates, the amount of data required, i.e., the number of time steps measured, should be considered. Fig. \ref{fig:ssizertot} shows the estimated total impedance of Subsystems I, II, and IV for different numbers of measurements with $\Delta t = 1 \; \mathrm{min}$ resolution, each with the actual total impedances shown as dashed lines. Observe that the estimate tends to approach the actual total impedance as the number of measurements increases. Two things stand out here: First, no continuous improvement occurs, which is due to the stochastic behavior of the underlying profiles. Second, the estimate is already in a reasonable range for small amounts of samples and, depending on the subsystem, obtaining accurate estimates after 10 minutes. These results show that good estimates of total impedance can be obtained for measured data over a single day and even for data over a few hours.

\begin{figure}[ht]
        \centering
        \includegraphics[width=\columnwidth]{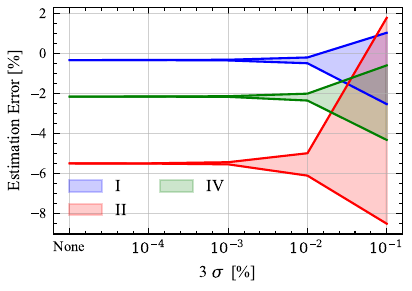}
        \caption{Effect of measurement noise on total estimated impedance.}
        \label{fig:noisertot}
    \end{figure}

    In addition to temporal resolution, the effect of measurement noise on the estimation of total impedances can be investigated. Thus, Fig. \ref{fig:noisertot} presents the results obtained by adding Gaussian noise to voltage and current measurements. Note that with no noise the results in Table \ref{tab:rtotesterrors} are obtained. Starting from a noise with $3\sigma = 10^{-3}$ the impact on the estimation quality becomes apparent. The filled area shows the results obtained from 100 Monte-Carlo simulations, highlighting that both increases and decreases in estimation error occur when adding noise to the system. Overall, for noise below $3\sigma = 0.1\%$, the estimation error is less than 10\%.
    
    \subsubsection{Cable Type Identification}
    \label{sec:resultscabletypeidentification}   
    When both estimated line lengths and estimated total impedance are available, the methodology presented in Section \ref{sec:cabletypeidentification} can be applied to estimate the employed line types. 
    
\begin{figure}[ht]
        \centering
        \includegraphics[width=\columnwidth]{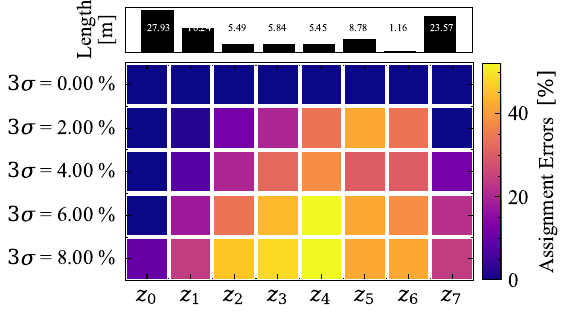}
        \caption{Cable identification error for segments in subsystem IV.}
        \label{fig:sub4endpoints}
    \end{figure}

    Thus, when the actual line lengths are available, as well as the actual total impedance is known, the line types can be estimated without error for all subsystems. However, as discussed previously, it the total impedance can not be determined without errors. Furthermore, errors in line lengths based on \gls{GIS} are likely. To deal with these errors, the estimation of line types must be robust. Fig. \ref{fig:sub4endpoints} shows the cable identification error for 50 Monte-Carlo simulations of Subsystem IV. Each column represents a segment of the subsystem and corresponds to the eight impedances that add up to the total impedance $\hat{z}_{\mathrm{tot}, \mathrm{IV}}$, and each row corresponds to the results obtained from 50 Monte-Carlo simulations with noisy input data, assuming Gaussian noise with standard deviation $\sigma$ for both the line lengths and total impedance $\hat{z}_{\mathrm{tot}, \mathrm{IV}}$ in Ohm. The color of each matrix element represents how often an incorrect cable type is selected for the respective segment. Observe that for more noise on the input data, an increase in cable type misclassifications occurs, resulting in false classifications in up to 50\% of the cases. Based on the length of each cable shown above the matrix, note that the misclassifications occur mainly for short cable segments, due to the small influence of the short segments on the total impedance, since a change of the cable type in the respective short segment does not impact the total impedance significantly. Therefore, when the estimation error of the total impedance is of the same order of magnitude as the impedance of individual cable segments, misclassifications of such segments inevitably occur.
    
    \subsubsection{Closed-Form Impedance Estimation}
    \label{sec:resultsclosedformimpedanceestimation}
\begin{figure}[ht]
        \centering
        \includegraphics[width=\columnwidth]{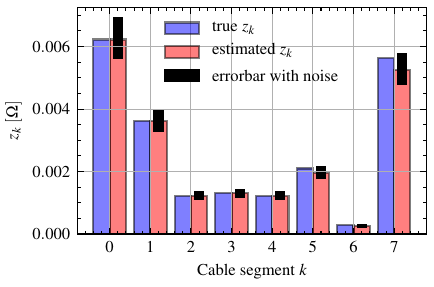}
        \caption{Estimation errors for Closed-form Identification for subsystem IV.}
        \label{fig:estimationErrorsClosed}
    \end{figure}

    The Closed-Form Impedance Estimation aims at solving the identification of parameters by making use of the lengths obtained from a \gls{GIS}-system, as described in Section \ref{sec:closedformimpedanceestimation}. Consequently, both the influences of errors in the line lengths and the measurement errors must be taken into account. In Fig. \ref{fig:estimationErrorsClosed} the estimates for Subsystem IV for all contained segments are shown, with the black bar representing the estimated range of the method with normally distributed noise on the segment lengths and current and voltage measurements, considering noise up to $3\sigma = 10 \%$. Observe that even for strong noise on the input data, reasonable results are obtained for the individual segments, which highlights the robustness of the method.    
    
    \subsection{Discussion of Methods}
    \label{sec:resultsoverall}
    Based on the available data, the procedure in Fig. \ref{fig:flowchart} can be performed in part or in its entirety, considering that steps can be executed independently from each other. Thus, the methods presented in Section \ref{sec:switchstateidentification} and Section \ref{sec:phaseidentification} can be used to identify the correct switch configuration, as well as to detect the phases, showing high robustness to errors in the measurement data, allowing to identify the correct topologies even under challenging circumstances.
    
    Two different methods for determining cable parameters and cable types are provided. The first method, presented in Sections \ref{sec:totalimpedanceidentification} and \ref{sec:cabletypeidentification}, divides the identification into two subproblems by first estimating the total impedance of the respective subsystem using regression methods, and then solving a \gls{BLP} to determine the cable types used. The second method presented solves a closed-form optimization problem as \gls{MIQP}. Both approaches lead to single-digit estimation errors in the considered scenarios. However, the results depend strongly on the quality of the estimation of the input data, leading to limitations in the quality of the results, especially for short cable segments.

    \begin{table}[ht]
        \centering
        \caption{Summary of obtained maximum estimation errors for tree topology and 50 Monte-Carlo runs}
        \begin{tabular}{|c|c|c|}
            \hline
            \textbf{Algorithm} & \textbf{Max. error [\%]}
            &\textbf{Max. error [\%]} \\
            & No Noise & \gls{AC} 0.5 S \\
            \hline
            Switch State Identification & 0.0 & 0.0 \\
            \hline
            Phase Identification & 0.0 & 0.0 \\
            \hline
            Total Impedance & 5.49 (II) & 34.94 (II) \\
            Estimation & & \\
            \hline
            Cable Type & 0.0 & 20.00 (I)\\
            Identification & & \\
            \hline
            Closed-Form & 6.67 (II) & 12.18 (II)\\
            Identification & & \\
            \hline
        \end{tabular}
        \label{tab:summary}
    \end{table}
    Table \ref{tab:summary} provides a summary of the worst-case results obtained for the test grid in tree configuration with the proposed methods. Note that, neglecting measurement noise, estimating the total impedance using regression leads to slightly better results compared to closed-form identification. However, when adding noise, the maximum estimation error increases to 34.94\% when using the regression method and to 12.18\% when using the closed-form identification, for which a noise on the line lengths of $3\sigma=10\%$ is assumed, when considering realistic measuring devices with \gls{AC} 0.5 S with 0.5\% maximum allowable measurement error. This shows that, given good estimates of the line lengths, a higher robustness can be achieved with the closed-form identification. Moreover, observe that the identification of the applied cable types is very sensitive to uncertainties in the input data, for several reasons: Thus, first the influence of two uncertainties becomes clear, from measurement noise on current and voltage and from assumed cable lengths. Second, the cable types used have very similar parameters, (e.g., 0.208 and 0.225 for $r'$), which results in classification errors. Lastly, by neglecting reactive powers, only an approximation of the underlying equations is considered.
    
    \section{Conclusions}
    \label{sec:conclusionoutlook}
    This paper presented methods for estimating topology and parameters in \gls{LV} electrical grids with limited available measurements. The proposed methodology deceomposed the identification into solvable subproblems, where individual methods can be applied independently if input data are available. The results showed that identification of switch states and phases of the measurements could be determined robustly for the given scenario. However, estimates of the total grid impedance using a data-driven regression algorithm presented errors due to the statistical nature of the problem, the large number of non-measurable quantities, and the simplifications made. 
    
    A method based on a \gls{MIQP} approach was also presented for identifying the cable types used, and a method for finding estimates of cable types in a closed-form optimization problem was formulated. Both approaches were compared regarding their performance, demonstrating their application and limitations with respect to the quality of the input data. Further investigations could be carried out combining measurements from smart meters with measurements obtained from grid equipment to develop methods that can estimate grid utilization, expansion requirements, and optimal schedules in scenarios with low sensor penetration. Additionally, exploring closed-form physical algorithms that can handle statistical uncertainties, such as Bayesian graphical models, could be considered.    
 
	\bibliographystyle{IEEEtran}
	\bibliography{bibliography}
\end{document}